\documentstyle[aps,epsfig,floats,preprint,amsfonts]{revtex}
\begin{document}
\draft 
\preprint{\vbox{{\hbox{\tt SOGANG-HEP 309/03 hep-th/0308002}}}}
\title{Exact soluble two-dimensional charged wormhole}
\author
{Ji Young Han\footnote{electronic address:opiumpop@string.sogang.ac.kr},
Won Tae Kim\footnote{electronic address:wtkim@mail.sogang.ac.kr}, and
  Hee Ju Yee\footnote{electronic address:elf@string.sogang.ac.kr}}
\address{Department of Physics and Basic Science Research Institute,\\
  Sogang University, C.P.O. Box 1142, Seoul 100-611, Korea}
\date{\today}
\maketitle
\begin{abstract}
We present an exactly soluble charged wormhole model in two
dimensions by adding infalling chiral fermions on the static wormhole.
The infalling energy due to the infalling charged matter requires the classical
back reaction of the geometry, which is solved by
taking into account of the nontrivial nonchiral exotic energy.
Finally, we obtain the exact expression 
for the size of the throat depending on the
total amount of the infalling net energy and discuss the interesting
transition from the AdS spacetime to the wormhole geometry.
\end{abstract}
\bigskip
\newpage
Two-dimensional
soluble gravitational models \cite{jt} are much simpler 
because of less degrees of 
freedom than the higher-dimensional counterparts in studying black
holes and cosmology. 
In fact, enormous works related to the intriguing gravitational issues
have been accomplished in this lower-dimensional regime
\cite{many,cghs,rst,hkl,no}.
Recently, wormhole solutions \cite{mt} in two-dimensional dilaton gravity
\cite{cghs,rst},   
have been obtained by adding an appropriate exotic matter which gives 
exact solvability \cite{hkl}.
Therefore, it may be natural to consider the exactly soluble charged
wormhole 
by adding dynamical
fermions, of course, the stability of the wormhole seems to be more or
less nontrivial. 

In this paper, we shall consider the infalling chiral fermions coupled
to the Abelian gauge field in the two-dimensional dilaton gravity. The Maxwell
kinetic term is slightly modified in order to exactly solve \cite{no}.
We consider left-handed chiral fermions coupled to the gauge
field, and the Abelian gauge field is suitably gauged away, which is
in fact only possible classically without the quantum gauge anomaly. 
We first derive a static charged wormhole solution in our model without the
infalling left-handed fermions and then the exact time dependent
charged wormhole solution is obtained by adding the fermions. 
The infalling matter spoils the
original wormhole structure due to the classical back reaction of the
geometry so that the left-right asymmetric exotic matter should
be considered to reproduce the exact wormhole geometry in later time. 

We now consider the Callan-Giddings-Harvey-Strominger(CGHS) 
dilaton gravity with chiral fermions coupled to
the U(1) gauge field, 
\begin{equation}
  \label{action}
  S = \frac{1}{2\pi} \int d^2x \sqrt{-g} \left[ e^{-2\phi}(R +
  4(\nabla \phi)^2+4\lambda^2)- \frac{e^{2\phi}}{g_A^2} F^2 - \sum_{j=1}^{N} i
  \bar\Psi_{j}\gamma^{\mu}(D_{\mu}-iA_{\mu})\Psi_{j} + \frac12(\nabla{\cal G})^2 \right],
\end{equation}
where $\phi $ and $\Psi_j = { 0  \choose \ \psi_j }$ are 
a dilaton field and the N left-handed chiral fermions, and
${\cal G}$ is a ghost scalar field whose energy density is negative.
And $g_A$ is the $U(1)$ electromagnetic gauge coupling constant
and $D_{\mu}$ is a covariant derivative. Note that the
gauge coupling  effectively depends on the dilaton field,
which is crucial to obtain the exact solution in later.
Defining  $ \Psi =$ exp$ (i\int {A_{\mu} dx^{\mu}}) X
$, $\bar\Psi = \bar X$exp$(-i {\int A_{\mu} dx^{\mu}}) $ in terms of
new variables $ X_j = {0 \choose \chi_{j}} $, and then in the conformal
gauge of $g_{+-} = -e^{2\rho}/2 $, $ g_{\pm\pm} = 0$, where
$ x^\pm = (x^0 \pm x^1) $, the metric equations of motion are given by
\begin{eqnarray}
 \label{metric eq pmpm}
 & &e^{-2\phi}(2\partial_{\pm}\partial_{\pm}\phi - 4\partial_{\pm}\rho\partial_{\pm}\phi)
 = T_{\pm \pm},\\
 \label{metric eq +-}
 & & e^{-2\phi}(2\partial_{+}\partial_{-}\phi -4\partial_{+}\phi\partial_{-}\phi-\lambda^2e^{2\rho})+\frac{2}{g_A^2}e^{2(\phi-\rho)}F_{+-}^2=0,
\end{eqnarray}
where the total energy-momentum tensors composed of the fermions and
the exotic field are $ T_{\pm \pm} = T_{\pm
  \pm}^f + T_{\pm \pm}^g$,
explicitly given as $ T_{++}^f =
 (i/4)\sum_{j=1}^{N}(\chi_{j}^{*}\partial_{+}\chi_{j}-(\partial_{+}\chi_{j}^{*})\chi_{j}),
 \quad T_{--}^f = 0$, and
$T_{\pm \pm}^g = -(\partial_{\pm}{\cal G})^{2}/2 $.
Then the Maxwell equations are also obtained as
\begin{equation}
 \label{Maxwell eq}
  \partial_{\pm}(e^{2(\phi-\rho)}F_{+-}) =J_\pm,
 \end{equation}
where the source currents are
$J_+=(g_A^2/16)\sum_{j=1}^{N}\chi_{j}^{*}\chi_{j}$
and $J_-=0$.
Next, the dilaton, fermions, and the ghost equations of motion are written as
\begin{eqnarray}
 \label{dilaton eq}
 & & -4\partial_{+}\partial_{-}\phi+4\partial_{+}\phi\partial_{-}\phi+2\partial_{+}\partial_{-}\rho+\lambda^2e^{2\rho}-\frac{2}{g_A^2}e^{4\phi-2\rho}F_{+-}^2=0,\\
 \label{fermion eq}
 & & \partial_{-}\chi_{j}=0,\\
 \label{ghost eq}
 & & \partial_{+}\partial_{-}{\cal G} = 0,
\end{eqnarray} 
respectively.
The fermions are free since we redefined them in terms of
new variables classically, which is impossible because of the chiral
anomaly in the quantized theory.

By using Eqs. (\ref{metric eq +-}) and (\ref{dilaton eq}), we get
the following simple relation, $\partial_{+}\partial_{-}(\rho-\phi)=0$,
which is key to the exact solubility of the charged wormhole,
and then the residual symmetry can be fixed by choosing $\rho = \phi $
corresponding to the Kruskal gauge.
Hence, the metric equations of motion can be simply written as
\begin{eqnarray}
 \label{constr}
 & & -\partial_{\pm}\partial_{\pm} e^{-2\phi} = T_{\pm \pm}^f + T_{\pm \pm}^g,\\
 \label{eq of mot}
 & & \partial_{+}\partial_{-}e^{-2\phi} + \lambda^2 -\frac{2}{g_A^2}F_{+-}^2 = 0,
\end{eqnarray}
and the Maxwell equations of motion (\ref{Maxwell eq}) becomes
$ \partial_{+}F_{+-} = J_+ $ and $ \partial_{-}F_{+-} = 0 $, 
and then the general solutions for 
Eqs. (\ref{fermion eq}) and (\ref{ghost eq}) are 
$ \chi_{j} = \chi_{j}(x^+)$ and $  {\cal G}(x^+,x^-) = {\cal
  G}_{+}(x^+) + {\cal G}_{-}(x^-)$.
These simple chiral fields are obtained with the help of the special
dilaton-gauge coupling in front of the Maxwell kinetic term in the
starting action (\ref{action}).

Before we derive the dynamical wormhole solution, we
exhibit a vacuum solution by assuming, $ X_j = {\cal G}=0 $, 
which is given as \cite{no}
\begin{equation}
 \label{csBH sol}
 F_{+-} = \frac{\bar Q}{2},~~ e^{-2\phi} = \frac{M}{\lambda} - {\tilde \lambda}^2 x^+x^-,
\end{equation}
where $M$ is a black hole mass and ${\tilde \lambda}$ is defined by 
$ {\tilde \lambda}^2 = \lambda^2 - 2C^2/g_A^2$ and $C$
is a constant corresponding to the background charge.
The structure of
spacetime is similar to that of the Schwarzschild black holes and
simple in contrast to the Reissner-Nordstr\"om black hole solutions in four
dimensions.

Next, we consider $ X_j = 0 $ and ${\cal  G} \neq 0 $ case 
with the special choice of 
$ {\cal G} = \sqrt{2} {\tilde\lambda}(x^+ - x^-)$,
the energy momentum tensor is then negative constant \cite{hkl},
\begin{equation}
 \label{csW T}
  T_{\pm\pm} = T_{\pm\pm}^g = -{\tilde \lambda}^2.
\end{equation}
Integrating Eq. (\ref{eq of mot}), we obtain the metric
$ e^{-2\phi} = - {\tilde \lambda}^{2}x^{+}x^{-} + a_{+}(x^{+}) +  a_{-}(x^{-})$
and the integration functions are determined
by the constraints (\ref{constr}) as $
 a_{\pm} = (1/2){\tilde \lambda}^2{x^{\pm}}^2 + B_{\pm}x^{\pm} +
 D_{\pm}$, where $B_{\pm}$ and $D_{\pm}$ are integration constants.
Choosing $B_{\pm} = 0$ and $D_+ + D_- = D$, the solutions is obtained as
\begin{equation}
 \label{csW sol}
  e^{-2\phi} = D + \frac{1}{2}{\tilde \lambda}^2(x^+ - x^-)^2,\\
\end{equation}
and the curvature scalar is calculated as
$ R = 4 {\tilde \lambda}^2 (D - (1/2){\tilde \lambda}^2 (x^+ -
    x^-)^2)/(D + (1/2){\tilde \lambda}^2 (x^+ - x^-)^2)$.
Note that for $ D>0 $, it is a traversable charged wormhole solution
with $ x^+ = x^- $ which is a condition of the coincidence
for the past and future horizons, whereas there is a naked
singularity curve along with $ D + (1/2){\tilde  
\lambda}^2 (x^+ - x^-)^2 = 0$ 
for $ D<0 $. Especially, for $ D=0 $, a constant negative curvature $ R
= -4{\tilde \lambda}^2 $ is given,
which is just anti-de Sitter(AdS) spacetime.
On the other hand, for ${\tilde \lambda}^2 < 0 $,
if $ D<0 $, the solution is ill-defined as seen from Eq. (\ref{csW
  sol}). Until now, we assumed positive $ {\tilde \lambda}^2 $ for simplicity.

We now study our model for the case of $X_j \neq 0$ and ${\cal  G}
\neq 0 $ in order to study the time dependent wormhole and its
maintenance. The infalling fermions carry the positive energy and the
electric charge simultaneously, which gives the charged wormhole solution.
However, in this dynamical situation in contrast to the static
wormhole, the initial wormhole geometry can not be maintained as time
goes on because of the classical back reaction of geometry. Therefore,
some additional corrections should be made to maintain the wormhole
geometry even at latest time. 

We now consider travelling charged
matter from our universe to the other by simply assuming $
\chi_{j}(x^+) \neq 0$ with the energy-momentum density expressed by a shock wave as    
\begin{equation}
 \label{shock:energy}
   T_{++}^f = \alpha\delta(x^+ - x_1^+),~~ T_{--}^f = 0,
\end{equation}
where $\alpha > 0$, and the electric charge density as
\begin{eqnarray}
 \label{shock:mw}
  J_+ = Q\delta(x^+ - x_1^+),~~ J_- = 0.
\end{eqnarray}
Note that we assumed the left-handed chiral fermions
since it is nontrivial compared to the left-right symmetric infalling
case, and both the charge density and the energy momentum tensor were
taken simply as the same shock-wave form.
Then from the Maxwell equation of motion (\ref{Maxwell eq}) and
the left-handed current (\ref{shock:mw}), 
\begin{equation}
 \label{charge}
  F_{+-} = Q\theta(x^+ - x_1^+) + \frac{\bar Q}{2},
\end{equation}
is given where $\bar Q/2$ is a background constant charge and
 $Q$ is added charge through the shock wave at $ x^+ = x_1^+ $. 

Next, integrating Eq. (\ref{eq of mot}) with Eq. (\ref{charge}), we obtain the
metric solution,
\begin{equation}
 \label{gensol:op}
  e^{-2\phi} = - {\tilde \lambda}^2 x^+ x^- + \frac{2\beta}{g_A^2}x^-(x^+ - x_1^+)\theta(x^+ - x_1^+) + a_+(x^+) + a_-(x^-),
\end{equation}
where $ \beta = Q^2 + Q\bar Q $. 
The shock wave source (\ref{shock:energy}) indicates 
irradiation of the wormhole from our
universe, which spoils the coincidence of the past and future event horizons.
So, in this infalling matter source, we should take into account some
corrected exotic matter source instead of the constant exotic
background Eq. (\ref{csW T}), so that we assume 
left-right asymmetric ghost energy-momentum tensors,
\begin{eqnarray}
 \label{ghost}
  \left.
   \begin{array}{ll}
    \displaystyle
    T_{++}^g =- {\tilde \lambda}^2 +
    \frac{2\beta}{g_A^2} x^-\delta(x^+ - x_0^+) - \alpha
    \delta(x^+ - x_0^+),\\
    \displaystyle
    T_{--}^g = -{\tilde \lambda}^2 -
    \frac{2\beta}{g_A^2} (x_1^- - x_0^-)\delta(x^- - x_0^-),
   \end{array}
  \right.
\end{eqnarray}
where $ x_1 > x_0$, which means that the negative energy source is
provided in advance before the fermions destroy 
the original static wormhole structure.
From Eqs. (\ref{metric eq pmpm}),
(\ref{shock:energy}) and (\ref{ghost}), we find the complete metric,
\begin{eqnarray}
 e^{-2\phi} & = & \frac12{\tilde \lambda}^2(x^+ - x^-)^2 -
 \frac{2\beta}{g_A^2}[x^-(x^+ - x_0)\theta(x^+ - x_0) -
 (x_1 - x_0)(x^- - x_0)\theta(x^- - x_0) \nonumber\\ 
  & & - x^-(x^+ - x_1)\theta(x^+ -
 x_1)] + \alpha(x^+ - x_0)\theta(x^+ - x_0) - \alpha(x^+ -
 x_1)\theta(x^+ - x_1) + D,
\end{eqnarray} 
and we assigned $x_0^+ = x_0^- = x_0$, $x_1^+ = x_1^- = x_1$ to satisfy
the wormhole condition in latest time $x^+ > x_1^+$ and $x^- > x_0^-$,
where the metric can be simplified by the 
desirable wormhole geometry for the limiting cases,
\begin{eqnarray}
\label{secsol}
  e^{-2\phi} = \left\{  
               \begin{array}{ll} 
               \displaystyle 
               \frac12{\tilde \lambda}^2(x^+ - x^-)^2 + D, & x^{\pm} <
               x_0,\\
               \displaystyle
               \frac12{\tilde \lambda}^2(x^+ - x^-)^2 
               + (\alpha -\frac{2\beta}{g_A^2}x_0)(x_1 -
               x_0)+ D, & x^+ > x_1 \textrm{ and } x^- >
               x_0.\end{array} \right.
\end{eqnarray}

Note that the size of the initial throat can be changed by
the given parameters, infalling energy density and electric charge.
To see how this infalling matter affects the throat, we calculate
the energy defined by \cite{cghs,rst}, 
\begin{equation}
  E = \int dx^+ x^+ T_{++} + \int dx^- x^- T_{--} ,
\end{equation}
which is explicitly calculated as
\begin{eqnarray}
  E = \left\{
      \begin{array}{ll}
      \displaystyle
      -\frac12 {\tilde \lambda}^2 ({x^+}^2 + {x^-}^2), & x^{\pm} < x_0,\\
      \displaystyle
      -\frac12 {\tilde \lambda}^2 ({x^+}^2 + {x^-}^2) +
      \frac{2\beta}{g_A^2} (x^-x_0 - x_0x_1 + {x_0}^2) + \alpha(x_1 - x_0), &
      x^+ > x_1 \textrm{ and }x^- > x_0. 
      \end{array} 
      \right.
\end{eqnarray}
Therefore, the change of the infalling energy is simply at $ x^-
  = 0 $ given by  
\begin{equation}
  \Delta E = (\alpha -\frac{2\beta}{g_A^2}x_0)(x_1 - x_0),
\end{equation}
and Eq. (\ref{secsol}) is rewritten in terms of $\Delta E$ such
as $ e^{-2\phi}= (1/2){\tilde \lambda}^2(x^+ - x^-)^2 + \Delta E +
D$, which implies the size of the throat is directly influenced by
the infalling energy.

We have obtained the exactly soluble charged wormhole solution in the
two-dimensional Maxwell-dilaton gravity. The infalling chiral
matter affects the static background geometry by adding 
the energy-momentum and electric charge. The classical 
back-reaction of the geometry has been properly taken into
account and the exact self-gravitating solutions are derived.
Of course, the corresponding exotic source should be adjusted
along with the infalling real matter, which has left-right asymmetric
form even if we only consider the left-handed chiral fermions.
On the other hand, unfortunately, 
we did not address the quantum back-reaction of the geometry due to
the conformal anomaly and the chiral anomaly, which might be more interesting in the
lower-dimensional quantum gravity. 

As a final comment,  as seen from the below of Eq. (\ref{csW sol}),
the curvature scalar is constant especially for $D=0$, which is nothing but the
AdS space time. So we may consider an interesting transition from the AdS spacetime to
the wormhole geometry by adding infalling matter. 
In fact, the AdS spacetime frequently appears in
near horizon physics of the charged black holes in a certain limit, so
the infalling real matter produces the wormhole geometry locally. In
our case, simply considering Eq. (\ref{secsol}) for $D=0$, the AdS
spacetime evolves into the wormhole geometry as far as 
$\alpha >{2\beta}x_0/{g_A^2}$.
Therefore, it will be interesting to study this kind of transition in
more detail.
  
\acknowledgments{This work was supported by the Korea Research
  Foundation Grant KRF-2002-042-C0010.}


\end{document}